\documentstyle[epsfig, 11pt] {article}
\begin{document}
\title {Singularity in classical and quantum Kepler Problem with
Weak Anisotropy}
\author {Zai-Qiao Bai and Wei-Mou Zheng \\
Institute of Theoretical Physics, Academia Sinica,
Beijing 100080, China }
\date{}
\maketitle
\baselineskip 24pt \vskip 5mm
\begin {minipage}{140mm}
\begin {center} {\bf Abstract} \end {center}
\baselineskip 24pt
Anisotropic Kepler problem is investigated by
perturbation method in both classical
and quantum mechanics.
In classical mechanics, due to the singularity
of the potential,  global diffusion in phase
space occurs at an arbitrarily small perturbation
parameter. In quantum mechanics, the singularity
induces a large transition amplitude between
quasi degenerate eigen states, which generically decays as
$\hbar$ in the semi-classical limit.
\end {minipage}

\newpage
The anisotropic Kelper problem (AKP)
is a 2-dimensional Hamiltonian system defined by
$$
H=\frac{1}{2}(p_x^2+p_y^2)-\frac{1}{\sqrt{\nu x^2+\mu y^2}}
\eqno (1)
$$
with $\mu>\nu>0$ and $\mu \nu=1$ (For a detailed discussion
of its dynamics, see \cite {Gut1} ). While the system
reduces to the Kepler problem when $\mu=1$,
the existence of chaos was rigorously
proved by Gutzwiller and Devaney in the case of
$\mu/\nu >9/8$ \cite{Gut2, Dev}.
The emergence of irregular dynamics dose not
follow the conventional KAM scenario due to
the singularity at the origin. In fact,
the extremely unstable motion in the vicinity of the
origin played a central role in Gutzwiller
and Devaney's proof. The dynamical implication
of the singularity when $\mu \rightarrow 1$ is, however,
still not clear.  When $\mu \approx 1$, (1) can be
rewritten as
$$
H=\frac{1}{2}(p_x^2+p_y^2)-\frac{1}{r}
-\epsilon\frac{\cos 2\theta}{r}=
H_{0}+\epsilon F,
\eqno{(2)}
$$
where $r=\sqrt{x^2+y^2}$ and $\theta=\tan^{-1}(y/x)$
while $\epsilon=(\mu-1)/2$ appears as a natural
perturbation parameter. In this letter we shall
study the classical and quantum AKP by perturbation
method. We hope this will provide a deeper understanding
of the quantum-classic correspondence in non-KAM system.

We begin with the classical mechanics.
Kepler problem is a maximally integrable system.
All its orbits are closed with period
$T=2\pi(-2E)^{-3/2}$, where $E=H_0<0$.
This global periodic motion will be destroyed
by an infinitely small perturbation. The first
order effect of a slight perturbation can be described
by the drift of closed orbit, i.e., the slow variation
of its parameters. Therefore, at this level of
approximation, we can consider the motion in the
space of all closed Kepler orbits (orbital space ).

The orbital space can be conveniently constructed
by taking advantage of the $so(3)$ dynamical symmetry.
Let
$$
  J_1=(\cos\theta-p_yJ_3)/\sqrt{-2E},~~J_2=(\sin\theta+p_xJ_3)/\sqrt{-2E},~~
  {\rm and}~~J_3=xp_y-yp_x.
\eqno (3)
$$
$(J_1,J_2,J_3)$ form a
$so(3)$ algebra, i.e., $\{J_i,J_j\}=\varepsilon_{ijk}J_k,i,j=1,2,3$
and $H_0=\frac{-1}{2(J_1^2+J_2^2+J_3^2)}.$
There is a one-to-one correspondence
between ${\mathcal J}=(J_1,J_2,J_3)$ and close Kepler orbits.
Specifically, write $(J_1,J_2,J_3)=\frac{1}{\sqrt{-2E}}
(\sin \alpha \cos\beta,\sin \alpha \sin\beta, \cos\alpha)$,
the corresponding Kepler orbit
in configuration space is defined as
$$
r=\frac{ J_3^2}{1-\sin \alpha \cos(\theta-\beta)}
\eqno (4)
$$
while the sign of $J_3$ determines its direction.
Therefore, the orbital space is coordinated by
$(J_1,J_2,J_3)$ ( or $(E, \alpha, \beta$)).

The variation of $J_i$ in one Kepler period is given by
$$
J_i(t+T)=J_i(t)+\epsilon T \{J_i,F_0({\mathcal J})\}+o(\epsilon),
~~~i=1,2,3,
\eqno(5)
$$
where $F_0$ is the time average of
$F$ over closed Kepler orbit,
$$
F_0=F_0({\mathcal J})=\frac{1}{T}\int _0^{2\pi}
F(r,\theta)\frac{r^2}{|J_3|}d\theta=
2E\frac {1-|\cos\alpha|}{1+|\cos\alpha|} \cos 2\beta.
\eqno (6)
$$
When $\epsilon\rightarrow 0$, Eq. (5) can be approximated by
a differential equation,
$$
\frac{d J_i}{d t}=\epsilon\{J_i,F_0({\mathcal J})\},~~~i=1,2,3.
\eqno (7)
$$
$\{H_0,F_0\}=0$ implies that the motion in the orbital space is
confined within a sphere ${\mathcal S}_E$ with $H_0=E=const.$.
Restricting $so(3)$ Poisson structure on ${\mathcal S}_E$ induces a
natural sympletic form
$
\omega_2=\frac{\sin \alpha}{\sqrt{-2E}} d\beta \wedge d\alpha,
$
which is, up to a constant, the ordinary area element.

As a two-dimensional Hamiltonian system, the dynamics on ${\mathcal S}_E$
can be easily determined by the contour chart of the
effective Hamiltonian $F_0$. There exist
six fixed points on the sphere.
The poles $(\alpha=0,\pi)$, which represent the two
circular Kepler orbits, are unstable while the
four on equator $(\alpha=\pi/2,\beta=0,\pm \pi/2,\pi)$,
which corresponds to the linear orbits on
the $x$ or $y$ axis, are stable.
The remaining orbits are either the
heteroclinic orbits ($\beta=\pm \pi/4,\pm 3\pi/4$)
that connect the poles or periodic orbits
surrounding one of the stable fixed points.
In other words, besides separatrixs, the classical
motion consists of four islands centered respectively
at the degenerate Kepler orbits on the $x$
and $y$ axis.

It should be pointed out that
the simple picture given by
perturbation analysis is not correct at
the vicinity of $\alpha=\pi/2$, i.e., the
collision orbits, where $\epsilon F$ is not bounded.
In fact, the first order derivative of $F_0$ is in general not
continuous at equator. This non-smoothness
manifests the non-perturbative
nature of the motion near the origin.
Noticing that all orbits on ${\mathcal S}_E$
except for the poles cross $\alpha=\pi/2$, we conclude that
the global dynamics of AKP cannot be described by
perturbation with respect to Kepler problem even at
the limit $\epsilon\rightarrow 0$.
Our numerical study show that the
collision orbits provide a passage-way
for global diffusion in the
orbital (and hence phase) space (Fig. 1).

Now we turn to quantum mechanics.
The $so(3)$ symmetry of quantum planar Kepler problem
is constructed in a way similar to its classical analog.
Specifically,
$$
\left\{\begin{array}{cl}
  J_1= & [\cos\theta-\frac{1}{2}(p_yJ_3+J_3p_y)]/\sqrt{-2H_0},
  \\\\
  J_2= & [\sin\theta+\frac{1}{2}(p_xJ_3+J_3p_x)]/\sqrt{-2H_0},\\\\
  J_3= & xp_y-yp_x,
\end{array}
\right.
\eqno (8)
$$
$[J_j,J_k]=i\varepsilon_{jkl}J_l$ and
$H_0=\frac{-1}{2(J^2+\frac{1}{4})}$ ($\hbar\equiv 1$).
Let $\{ |n,m>: m\leq |n|, n=0,1,... \}$ be
the standard $so(3)$ orthonormal set, i.e.,
$J_3 |n,m>=m|n,m> $ and $J^2|n,m>=n(n+1)|n,m>$.
$|n,m>$ is an eigenstate of $H_0$ with energy
$E=E_n=\frac{-1}{2(n+\frac{1}{2})^2}$ and wave function
in coordinate representation given by
$$
\Psi_{n,m}(r,\theta)=(-1)^{m}\frac{1}{|2m|!}\sqrt{\frac{(n+|m|)!}
{(n-|m|)!}}\frac{4 s^{|m|}}{(2n+1)^{3/2}}
e^{-\frac{1}{2}s }F(-n+|m|,2|m|+1,s)
\frac{e^{im\theta}}{\sqrt{2\pi}}
\eqno(9)
$$
where $s=4r/(2n+1)$ and $F(a,b,x)$ is the confluent hypergeometric
function.

The first order effect of a slight perturbation
is the mixing of states with definite $n$. In the
interaction representation, the long-time evolution
when $\epsilon \rightarrow 0$ is given by an effective
Sh$\ddot o$rdinger equation
$$
i \frac{d}{d\tau}\Psi=\overline {F} \Psi
\eqno (10)
$$
where $\overline {F}$ is the restriction of $F$ in eigen spaces
of $H_0$, i.e.,
$$
<n',m'|\overline {F}|n,m>=\delta_{n,n'}<n',m'|F|n,m>.
\eqno (11)
$$
Eq. (10) can be regarded as the quantum counterpart of
Eq. (6) and hence $F_0$ is the classical correspondence of
$\overline{F}$. The matrix elements of $\overline {F}$
is readily evaluated in coordinate representation,
$$
<n,m'|\overline{F}|n,m>=
E_n\sqrt{\frac{(n+m_1)!(n-m_1)!}{(n+m_2)!(n-m_2)!}}\delta_{|m'-m|,2},
\eqno(12)
$$
where $m_1=\min \{|m|,|m'|\}$ and $m_2=\max \{|m|,|m'|\}$.

By diagonalizing $\overline{F}$ in each
$2n+1$-dimensional subspace, we can study
the classical-quantum correspondence
in the framework of perturbation theory.
Notice that $\exp(i\frac{\pi}{2} J_3) \overline {F}+
\overline {F}\exp(i\frac{\pi}{2} J_3)=0$,
the spectrum of $\overline {F}$ is symmetric with respect to $0$.
We shall focus on the positive part, which corresponds to
the classical islands centered at fixed
points $(\alpha,\beta)=(\pi/2,\pi/2)$
and $(\pi/2,3\pi/2)$.

The classical orbits centered at fixed point $(\pi/2,\pi/2)$
contribute to the spectrum of $\overline {F}$
according to the semi-classical quantization rule,
$$
\frac{1}{2\pi\hbar}\int_{\sigma(\lambda_k)}
\omega_2=k+\mu_s/4, ~~~k=0,1,...
\eqno(13)
$$
where $\sigma(\lambda)\subset S_{E}$ is the region
enclosed by orbit with $F_0(E,\alpha,\beta)=\lambda$
and the Maslov index $\mu_s=2$.
For convenience, we rescale $E=-\frac{1}{2}$ and
$\hbar=\frac{2}{2n+1}$ so that $|F_0|\le 1$ and
Eq. (13) yields
$$
S(n,\lambda_k)\equiv \frac{n+\frac{1}{2}}{\pi}(\cos^{-1}\lambda_k+
\frac{2\lambda_k\ln \lambda_k}{\sqrt{1-\lambda_k^2}})=k+\frac{1}{2}.
\eqno(14)
$$
The orbits surrounding point $(\pi/2,3\pi/2)$ gives
the same contribution so that the semi-classical
spectrum consists of 2-fold degeneracies.
Eq. (14) is a good approximation of the exact spectrum (Table 1).

The degeneracy predicted by the semi-classical rule
is not exact due to quantum tunneling between the
two quasi static states each of them classically
corresponds to one periodic orbit. The splitting of
spectrum $(\Delta \lambda)$
is connected with the transition amplitude accumulated
in one Kepler period ($A_T$) by
$A_T=\epsilon \Delta \lambda T/2\hbar$.
In conventional quantum system, such as a particle confined in
double-well potential, $\Delta\lambda$ decays as
$\sim \hbar^{\gamma}\exp(-S/\hbar)$ when
$\hbar\rightarrow 0$, which vanishes faster
than any power of $\hbar$.
Fig. 2 shows $\Delta\lambda_{sc}\equiv\Delta\lambda/\hbar^2$ for
$n=100$ and $1000$, from which we can see that
$\Delta\lambda\sim \hbar^2$ (or $A_T\sim \hbar$) in the
semi-classical limit.
As a characteristic phenomenon in
in non-smooth system\cite{Bai}, this power-law reflects the
non-smoothness in $F_0$ and hence the singularity in AKP.

In summary, we have studied in this letter the Anisotropic
Kepler problem by perturbation method. By using the dynamical
symmetry of the unperturbed system, we obtained
an effective Hamiltonian in both classical and
quantum mechanics.
We shown that the long-time evolution of
most of the classical orbits is non-perturbative
due to the singularity at origin and
global diffusion in phase space  occurs even at
an arbitrarily small perturbation parameter.
In quantum mechanics, we shown that
the characteristic phenomenon attributed
the singularity is the the power-law $\hbar$-dependence of
tunneling amplitude between states
which classically correspond to
separated periodic orbits.

\newpage
\begin {center}
Table

\begin{tabular}{|c|c|c|c|c|}
 \hline
 $k$& $\lambda$ & $S(n,\lambda)$ & $\lambda$ &$S(n,\lambda)$ \\
 \hline
 0& 0.6883 & 0.34 & 0.6512 & 0.42 \\
 1& 0.3335 & 1.52 & 0.3264 & 1.55 \\
 2& 0.1944 & 2.43 & 0.1875 & 2.48 \\
 3& 0.0898 & 3.50 & 0.0856 & 3.55 \\
 4& 0.0307 & 4.43 & 0.0293 & 4.46 \\
  \hline
\end{tabular}

Tab.1 Spectrum of $\overline {F}$ at $n=10$.
\end{center}

\newpage
\begin{figure} [htbp]
\vspace{-9cm}
  \epsfig{file=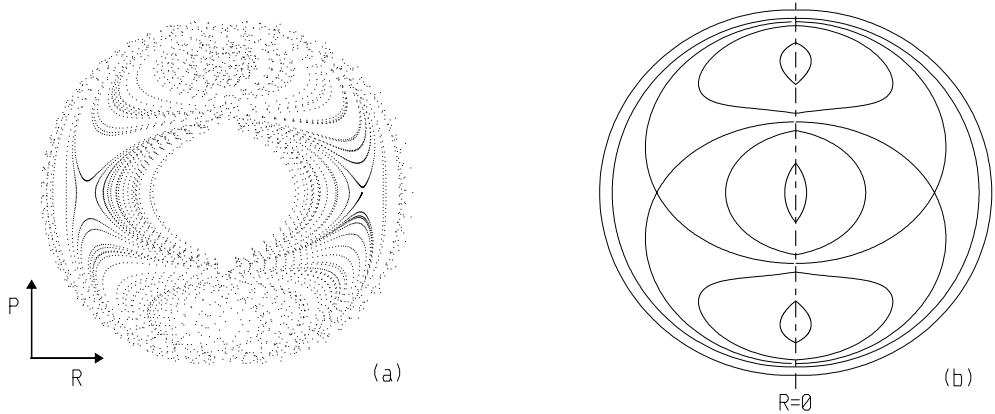,width=16cm}
\vspace{-8cm}
\caption{(a) Numerical calculation of one orbit at $\mu=1.04$.
$5000$ intersections are plotted on the Poincar$\acute{e}$
surface of section defined at $\theta=0$.
with coordinate $(R,P)={\rm sign} (p_y)(\sqrt{r},\sqrt{r}p_x)$,
comparing with the regular structure implied by the first-order
perturbation (b).
}
\end{figure}

\begin{figure} [htbp]
\vspace{-8cm}
 \epsfig{file=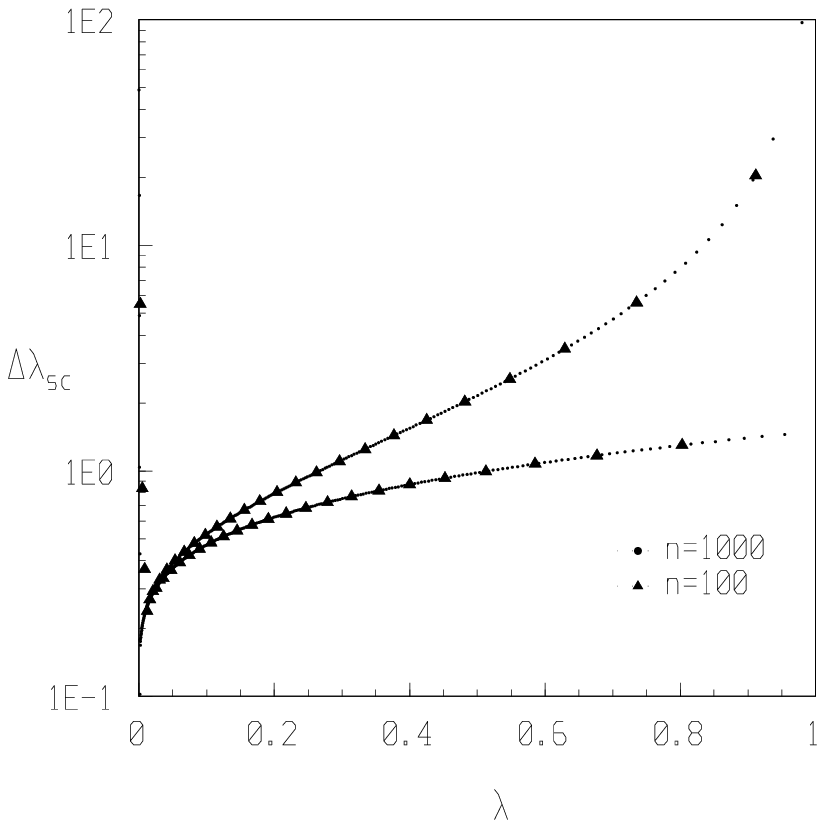,width=16cm}
\vspace{-8cm}
\caption{ Scaled spectrum splitting ($\Delta \lambda_{sc}$) for $n=100$ and
$1000$. Except for few points near $\lambda=0$,
$(\lambda,\Delta \lambda_{sc})$ is approximately located
at two curves (determined by whether $k$ is even or odd)
irrespective of $n$.}
\end{figure}

\begin{thebibliography}{9}
\bibitem{Gut1}  M. C. Gutzwiller, Chaos in Classical and Quantum Mechanics
              (Springer, New York, 1990).
\bibitem{Gut2} M. C. Gutzwiller, J. Math. Phys. 18, 806 (1977).
\bibitem{Dev}  R. L. Devaney, Invent. Math. 45, 221 (1978).
\bibitem{Bai} Z. Q. Bai preprint (quant-ph/0110126)

\end{thebibliography}
\end {document}